\documentclass[aps,pre,twocolumn,superscriptaddress,showpacs]{revtex4}
\usepackage{amsfonts,amssymb,amsmath,latexsym,epsfig,wasysym} 

\begin{document}

\title{Chaotic Explosions}

\author{Eduardo G. Altmann}
\affiliation{Max Planck Institute for the Physics of Complex Systems, 01187 Dresden,
  Germany}
\author{Jefferson S. E. Portela}
\affiliation{Max Planck Institute for the Physics of Complex Systems, 01187 Dresden,
  Germany}
\author{Tam\'as T\'el}
\affiliation{Institute for Theoretical Physics - MTA-ELTE Theoretical Physics Research  Group, E\"otv\"os University, Budapest, H--1117, Hungary}

\begin{abstract}
We investigate chaotic dynamical systems for which the intensity of trajectories might grow
unlimited in time. We show that (i) the intensity grows  exponentially in time and is
distributed spatially according to a fractal measure with an information
dimension smaller than that of the phase space, (ii) 
such exploding cases can be described by an operator formalism similar to the one applied to chaotic systems with absorption (decaying
intensities), but (iii) the invariant 
quantities characterizing explosion and absorption are typically not directly related to
each other, e.g., the decay rate and fractal dimensions of absorbing
maps typically differ from the ones computed in the corresponding inverse (exploding) maps. We illustrate our general results through
numerical simulation in the cardioid billiard mimicking a lasing optical cavity, and
through analytical calculations in the baker map.
\end{abstract}
\pacs{05.45.-a,05.45.Df,42.55.Sa}

\maketitle

\section{Introduction}

Fractality is a signature of chaos appearing in strange attractors and in the invariant
sets of open dynamical systems~\cite{Ott-book,LaiTel-book,GaspardBook}. Here we are
  interested in systems in which trajectories have associated to them a time-varying {\it
    intensity}. In a recent work~\cite{PRL} we
showed that fractality in chaotic 
systems in which the intensity of trajectories decays due to {\em absorption} can be described by an operator formalism and that absorption leads to a multi-fractal spectrum of the decaying state.  In this
work we investigate the dynamical and fractal properties of systems containing {\em gain} (i.e., 
negative absorption). In systems with gain the energy or intensity of trajectories
increases in time, e.g., the intensity of a ray grows while it is in a gain medium or is multiplied by a factor  larger than one when reflected on a wall.

\begin{figure*}[!bt]
\begin{center}
\includegraphics[width=0.8\columnwidth]{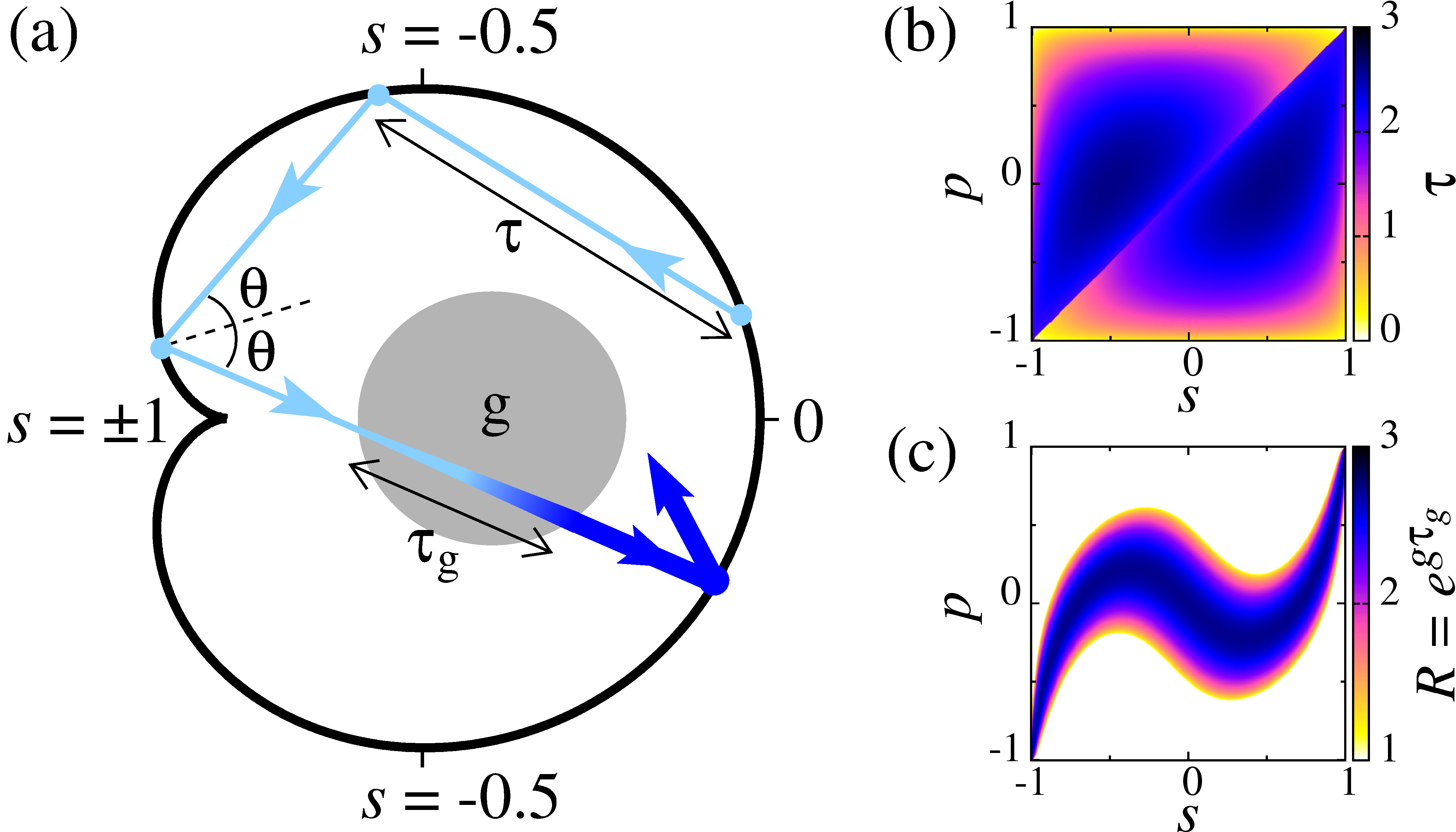}\hspace{0.1\columnwidth}\includegraphics[width=0.95\columnwidth]{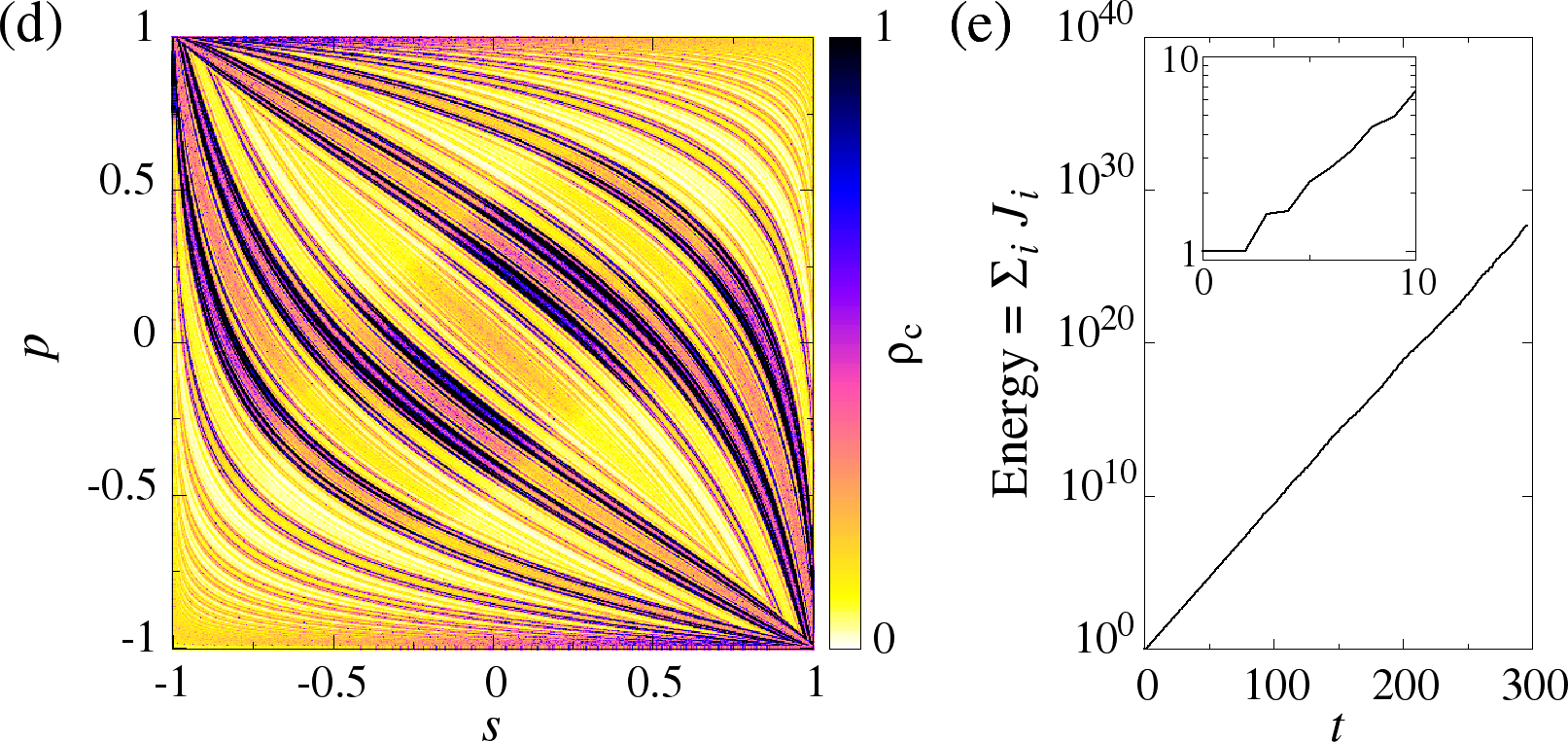}
\end{center}
\caption{Explosion in a fully chaotic billiard.
(a) Cardioid billiard, whose boundary in polar coordinates is
$r(\phi)=1+\cos(\phi)$ with $\phi \in [-\pi,\pi]$~\cite{Robnik1983}. The gain region (gray, marked by g) is a disc of radius $0.5$ in the middle of the billiard.
(b) Collision time distribution $\tau(\vec{x})$ in the cardioid billiard (velocity modulus is unity). Birkhoff coordinates $\vec{x}=(s, p=\sin \theta)$ are used, where $s$ is the arc length along the boundary and $\theta$ is the collision angle.
(c) Reflection coefficient distribution: $R(\vec{x})=e^{g \tau_g}$ with $g=1$ and $\tau_g$ given by
the length of the intersection of the ray with the gain region. 
(d) Time-independent density~$\rho_c$ in the phase space.
(e) Time-dependence of the intensity integrated over the phase space.  The explosion rate
$\kappa\approx0.215$ is  the slope of the curve (note the log-scale). Inset shows the
non-exponential behavior for short times. }
\label{fig1}\label{fig.bill}
\end{figure*}

Optical microcavities provide a representative physical system of the general
dynamical-systems picture described above.  The formalism of open
chaotic systems has been extensively used to describe lasing properties of
two-dimensional optical cavities~\cite{Harayama2011,RMP}. The success of
this approach relies on the use of long-living ray trajectories to describe the lasing
modes. This is justified because lasing modes are induced by the gain medium present in optical cavities and
only long-living trajectories are able to profit from this gain. The relevance
of gain led to specific investigations of its role in experiments~\cite{Shinohara2011} and
wave simulations~\cite{Martina2013}, but we are not
aware of ray simulations which have explicitly included gain. This is a
crucial issue especially when the gain is not uniformly distributed in the cavity, as in the experiments of Ref.~\cite{Shinohara2011}.

We consider chaotic billiards with gain as models of optical microcavities. In
Fig.~\ref{fig1} we show simulations of trajectories that bounce elastically, but whose intensities
increase exponentially in time with rate $g$ while passing through a gain region (the gray
disc in Fig.~\ref{fig1}a).
For long times, the total intensity grows exponentially in time and the
spatial distribution (obtained for any $t$ normalizing over the phase space) approaches a
fractal density $\rho_c$.  We call this phenomenon a {\it chaotic explosion}.

In this Letter we show that chaotic systems with gain can be treated with the same
 formalism of systems with absorption but that the properties of these two classes of systems
 cannot be trivially related to each other. We obtain general results for the fractality
 and for the inverse map of systems with gain/absorption, which are illustrated
 analytically in the baker map and numerically in the  cardioid  billiard. For optical
 microcavities, our results  show how gain can be introduced in the ray description and
 how it affects the far-field emission, demonstrating also in the ray description that
 lasing is not determined by the shape of the (passive) cavity alone.

\section{True-time maps with gain}\label{sec.formalism}

We consider an extended map which includes, besides a usual map $f$, the true physical time $t_n$ and the ray
intensity $J_n$ at the $n$-th intersection  $\vec{x}_{n}$ with a Poincar\'e section  as~\cite{Gaspard1996,Kaufmann2001}
\begin{equation}\label{eq.extended}
f_{\text{ext}}:\left\{\begin{array}{lll}
\vec{x}_{n+1} & =f(\vec{x}_{n}),  \\
t_{n+1} & = t_n + \tau(\vec{x}_{n}),\\
J_{n+1} & = J_n R(\vec{x}_{n}),
\end{array}
\right.
\end{equation}
where the {\it return time} $\tau(\vec{x})\ge0$, chosen as the time between intersections $\vec{x}=\vec{x}_n$
  and $\vec{x}_{n+1}=\vec{x}'\equiv f(\vec{x})$  (for billiards, $\tau$ is the collision
  time between two consecutive bounces with the wall), 
and the reflection coefficient
$R(\vec{x})$ are known functions of the coordinate $\vec{x}$ on the Poincar\'e section. 
Cases in which gain occurs continuously in time (not only at
  the intersection with the Poincar\'e section) correspond to a reflection coefficient $R(\vec{x}) = e^{g \tau_g(\vec{x})}$,  
where $g$ is the gain rate and
  $\tau_g$ is the time spent in the gain region ($\tau_g=\tau$ if gain is uniform in the
  billiard table).
Explosion occurs if $R(\vec{x})>1$ for a sufficiently large region of
$\vec{x}$.

We are  interested in the density~$\rho(\vec{x},t)$ (i.e., the collective intensity of an
ensemble of trajectories) at time $t$ in $\vec{x}$. Here we consider the class of
(ergodic and chaotic) maps $f(\vec{x}_{n})$ for which we show that for any smooth initial
$\rho(\vec{x},t=0)$ one observes for long times 
\begin{equation}
\rho(\vec{x},t) \sim \rho_c(\vec{x})e^{\kappa t},
\label{eq.rhos}
\end{equation}
where $\kappa$ is the temporal rate of the total energy change (an explosion rate for $\kappa>0$)
and is independent of $\rho(\vec{x},t=0)$. 

We can expect that
$\rho_c(\vec{x})$ of (\ref{eq.rhos}) is the attracting
density of an iteration scheme for $\rho(\vec{x})$ of the
extended map~(\ref{eq.extended}). With a compensation factor $e^{-\kappa \tau(\vec{x})}$
per iteration, 
this scheme evolves a density $\rho_{n}(\vec{x})$ at discrete time $n$
into $\rho_{n+1}(\vec{x})$ at the next intersection with the Poincar\'e surface of section as 
\begin{equation}
 \rho_{n+1}(\vec{x}') =  \sum_{\vec{x}\in {f}^{-1}({\vec{x}')}}  e^{-\kappa \tau(\vec{x})}
                \frac{R(\vec{x}) \rho_{n}(\vec{x})}{\mid \mathcal{D}_f(\vec{x})\mid},
\label{eq.FP}
\end{equation}
where $\mathcal{D}_f$ represents the Jacobian of the Poincar\'e map $f$.
 In the special case of invertible area-preserving dynamics (as in the billiard of
 Fig.~\ref{fig.bill}), $\mathcal{D}_f(\vec{x})=1$ and there is no sum in (\ref{eq.FP}) (map $f$ has a single preimage).

In any extended map, there exists one $\kappa$ -- the one appearing in Eq.~(\ref{eq.rhos})
-- for which $\rho_c(\vec{x})$  arises as the limiting ($n \rightarrow \infty$)
distribution of $\rho_{n}(\vec{x})$ iterated by scheme (\ref{eq.FP}).  The right-hand-side
of Eq.~(\ref{eq.FP}), with the proper $\kappa$, is an operator
with largest eigenvalue unity and $\rho_c$ as the corresponding 
eigenfunction. The second-largest eigenvalue controls the convergence of a smooth
  initial density $\rho_0$ to $\rho_c$. In agreement with the physical
  picture, we see that for extended maps $f_{\text{ext}}$ there are three different
  factors contributing to the  density $\rho$: reflectivity $R$, return times
  $\tau$, and stretching of phase space volume $\mathcal{D}_f$. The rate $\kappa$ follows
  from the constraint that the compensated intensity neither increases nor decreases for large
  $n$~\cite{LaiTel-book}. 

Comparing with our previous results~\cite{PRL,RMP}, we see that  (\ref{eq.FP}) is an
extension to cases with $R>1$ of the  
Frobenius-Perron operator considered previously only in systems with absorption ($R\le1$
for any $\vec{x}$).  The explosion rate $\kappa$ plays the role of a negative escape
rate and the attracting limit distribution $\rho_c$ is the conditionally invariant density~\cite{RMP} (also known as the steady probability distribution in the optics
community~\cite{Lee2004,Ryu2006}). 
By integrating, for $n \rightarrow \infty$,  (\ref{eq.FP}) over all $\vec{x}'$, the left
hand side is unity due to normalization of $\rho_{n}(\vec{x})$,
while the right hand side is an average taken with respect to $\rho_c$. This yields (see
also Ref.~\cite{PRL})
%
$\langle  e^{-\kappa \tau} R \rangle_{c}=1.$
%
For chaotic explosions this means that  typically $R>1$, but its reduced value 
$e^{-\kappa \tau} R$ taken with the proper explosion rate averages out to unity if
the average is taken with the $\rho_c$ associated to $\kappa$.

\section{General Features}\label{sec.questions}

It is remarkable that the formalism of transient chaos~\cite{GaspardBook,LaiTel-book} can be
applied with slight modifications to describe chaotic 
explosions. Below we use this formalism to explore the most interesting effects of gain
($R>1$).

\subsection{Fractality}

\begin{figure}[!bt]
\begin{center}
\includegraphics[width=0.5\columnwidth]{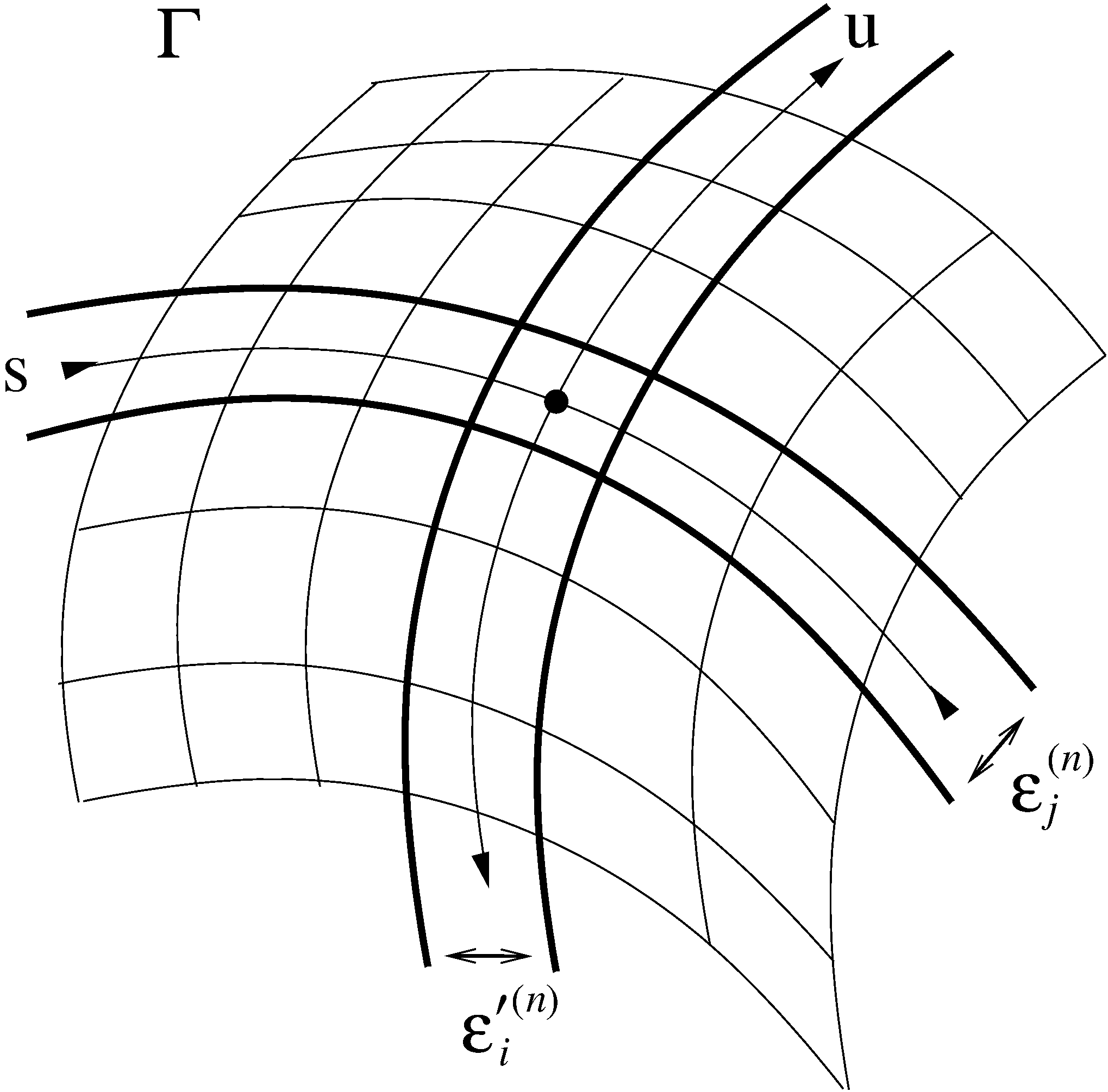}
\end{center}
\caption{Schematic diagram of the phase-space partitioning
around a hyperbolic chaotic set obtained taking the $n\gg1$ fold image and preimage
of the region of interest $\Gamma$ with respect to the (usual) map $f$. These define narrow strips and columns   
overlapping with branches of the stable (s) and unstable (u) manifold, respectively. 
The emphasized strip and column belong to a $n$-cycle point ($\bullet$)
on the chaotic set.}
\label{fig.hyperbolic}
\end{figure}

We first derive general relations between the fractality of  $\rho_c$ -- the
eigenfunctions of Eq.~(\ref{eq.FP}) -- and the  distributions~$(\tau(\vec{x}),R(\vec{x}))$  characterizing the
extended map $f_{\text{ext}}$. Let us consider the map $f$  in (\ref{eq.extended}) to be
invertible and two-dimensional.  We are interested in the dynamics
within a region of interest $\Gamma$ only,  which has a size of order unity in
appropriately chosen units.  Considering the $n$-th image and preimage of $\Gamma$ taken with the map $f$, we find for $n \gg 1$ a set
of narrow ``columns'' in the unstable direction and narrow ``strips'' along the stable
directions, as illustrated in Fig.~\ref{fig.hyperbolic}. They are good approximants of the unstable and stable foliation of the chaotic
set (a chaotic sea or an attractor in closed systems, or a chaotic saddle in open ones) underlying the dynamics.  
Each of them contains an element of an $n$-cycle~\cite{Ott-book,LaiTel-book}, i.e., a
point which is mapped by $f$ onto itself 
after $n$ iterations.

Let us focus on such an $n$-cycle point at the intersection of strip $j$ and column $i$
(drawn with bold lines in Fig.~\ref{fig.hyperbolic}).
Due to the permanent contraction in the stable direction, the width of column $i$ is approximately 
$\varepsilon_i'^{(n)}=e^{{\lambda'}_i^{(n)} n}$ where 
${\lambda'}_i^{(n)}<0$ is the contracting Lyapunov exponent around  the cycle point over discrete time $n$.
Similarly, the height $\varepsilon_j^{(n)}$ of strip $j$ is
 $\varepsilon_j^{(n)}=e^{-\lambda_j^{(n)} n}$, where $\lambda_j^{(n)}>0$ denotes the corresponding positive Lyapunov exponent.
By construction, points starting in strip $j$ spend the dominant part of their lifetime $n$ in the close {\em vicinity of the hyperbolic
  cycle} in question (which belongs to the chaotic set). Therefore, for large $n$ nearly all points in the strip are subjected to an average 
stretching factor $e^{\lambda_j^{(n)} n}$ (in the unstable direction). In an extended map $f_{\text{ext}}$,
obtained from $f$ through Eq.~(\ref{eq.extended}), these points
experience an average collision time $\tau^{(n)}_j=\sum_{k=1}^n \tau_k/n$ and an average
gain/reflection coefficient $R^{(n)}_j=(\Pi_{k=1}^n R_k)^{1/n}$  while being around the $n$-cycle, where  
$\tau_k$ and $R_k$ denote the collision time and reflection coefficient, respectively, belonging to element $k$ of
the $n$-cycle.

In the spirit of operator (\ref{eq.FP}) valid for $f_{\text{ext}}$, the 
density on the images of strip $j$ is the density of strip $j$ multiplied (in each iteration) by a factor
$R_k e^{-\kappa \tau_k}$. Starting with an initial unit density, the area  
$\varepsilon_j^{(n)} \times 1$ of strip $j$ should be multiplied in $f_{\text{ext}}$
by a factor $e^{-\kappa \tau^{(n)}_j n + n \ln{R^{(n)}_j}}$ by the end of $n$ iterations. 
Since the $n$-th image of strip $j$ with respect to $f$ is column $i$, by construction, the measure accumulating on
column $i$ is 
\begin{equation}
\mu^{(n)}_i=e^{-\kappa \tau^{(n)}_j n + n \ln{R^{(n)}_j}} \varepsilon^{(n)}_j = 
e^{\left(\ln{R^{(n)}_j}-\kappa \tau^{(n)}_j -\lambda_j^{(n)}\right) n }.        
\label{eq.mui}
\end{equation}     
The existence of a time-independent measure implies 
that with the 
proper value of $\kappa$ we have $\sum_i \mu^{(n)}_i=1$. 

In the spirit of dynamical systems theory, we associate with strip $j$ the measure
that its points represent after $n$ steps. Therefore the measure $\mu^{(n)}_j$ of strip $j$ is  
\begin{equation}
\mu^{(n)}_j=\mu^{(n)}_i.
\label{eq.muimuj}
\end{equation}
The relation expresses that $f_{\text{ext}}$ maps the measures of the stable and unstable
directions  into each other.

We now focus on fractal properties of this measure. 
Systems described by closed maps $f$ with gain have a trivial fractal dimension
$D_0$ 
equal to the phase space dimension. Their fractality requires thus the computation of the 
generalized dimensions~\cite{Ott-book,LaiTel-book}
\begin{equation}\label{eq.dq}
D_q=\frac{1}{1-q} \lim_{\varepsilon \rightarrow 0} \frac{\ln \sum_k \mu_k^q}{\ln 1/\varepsilon}
\end{equation}
where $\mu_k$ is the measure of the $k$-th box in a coverage with a uniform grid of box
size $\varepsilon$, and the sum is over  non-empty boxes. 

A general result can be obtained for the information dimension $D_1$. For a general
one-dimensional  distribution containing measures $\mu_{\alpha}$ in intervals of different sizes
$\varepsilon_{\alpha}$ we have for small $\varepsilon_{\alpha}$~\cite{Ott-book,LaiTel-book}:
$D_1 = \sum_{\alpha} \mu_{\alpha} \ln{\mu_{\alpha}}/\sum_{\alpha} \mu_{\alpha} \ln{\varepsilon_{\alpha}}$.
Now, take a section of strip $j$ along the unstable foliation (u), and associate
the measure of the strip $\mu_j^{(n)}$ to the interval size  $\varepsilon_j^{(n)}$. In the limit 
$n\gg1$ the sizes are small, and 
substituting Eq.~(\ref{eq.mui}), (\ref{eq.muimuj}) in the $D_1$ formula above, we
obtain the partial information dimension $D^{(1)}_1$ along the unstable direction as 
\begin{equation} \label{eq:D1}
D^{(1)}_1 =1 + \frac{\kappa \bar{\tau} -\overline{\ln R}}{\bar{\lambda}}.
\end{equation}   
The averages denoted by overbars are taken over the chaotic set (e.g., $\bar{\lambda}$ is
  the positive Lyapunov exponent on the saddle). This is so because, as argued above, the trajectory effectively experiences collision times $\tau$,  reflection
coefficients~$R$, and local Lyapunov exponents $\lambda$ (of the usual map $f$)  close to
an unstable cycle (which belongs to the chaotic set). 
Quantities characterizing the map ($\bar{\lambda}$), the collision times
($\bar{\tau}$), and the gain ($\overline{\ln R}$) determine a fractal property
($D_1$) via the simple relation~(\ref{eq:D1}). 
Fractality is nontrivial if $D^{(1)}_1<1$, implying $\kappa \bar{\tau} <\overline{\ln R}$ (for a positive rate $\kappa$),
 i.e.,
$R>1$ in a sufficiently extended region.   
 
Repeating the calculation along a section parallel to the 
stable foliation (s) we find $D^{(2)}_1$, the 
partial information
dimension along the stable direction. 
Since the measures are the same (see (\ref{eq.muimuj})), the difference follows from the
sizes which are 
$\varepsilon_i^{'(n)}$ now, and we find
\begin{equation}
D^{(2)}_1 \mid \bar{\lambda'} \mid = D^{(1)}_1 \bar{\lambda}.
\label{eq.D2}
\end{equation} 
The overall information dimension of the chaotic set  with the measure of $f_{\text{ext}}$
is $D_1=D^{(1)}_1+D^{(2)}_1$.
The derivation above holds for any value of $R$ and therefore generalizes the results of
Ref.~\cite{PRL}, where we conjectured Eq.~(\ref{eq:D1}) (with a negative $\kappa$) for strictly absorbing cases
($R\le 1$). 
The case of usual maps ($\tau=1, R=1$) follows as a limit: if the map is closed (no loss
in any sense), $\kappa=0$, and $D_1=1+\bar{\lambda}/\mid \bar{\lambda'} \mid$, i.e. we recover the Kaplan-Yorke
formula~\cite{Ott-book} ($D_1=2$ for area preserving maps); if the map is open
(trajectories escape), $\kappa<0$, and Eqs.~(\ref{eq:D1}) and  (\ref{eq.D2}) go over into the
Kantz-Grassberger formulas~\cite{GaspardBook,LaiTel-book}.      

\subsection{Inverse map}\label{sec.inverse}

Besides the physical motivation to study gain, a natural question that can only be
answered considering both $R>1$ and $R<1$ is the one of the 
inverse of the extended map defined in Eq.~(\ref{eq.extended}), which we denote by
$f^{\text{inv}}_{\text{ext}}$ and define implicitly by 
$f_{\text{ext}} \circ f^{\text{inv}}_{\text{ext}}=I,$
%
where $I$ is the identity. If $f_{\text{ext}}$ has $R>1$, then
$f^{\text{inv}}_{\text{ext}}$ should compensate this with $R<1$.  

For $f^{\text{inv}}_{\text{ext}}$ to exist, the usual map $f$ has to be invertible (i.e., $f$ must have a single pre-image). Consider the case in
which the dynamics is defined in a bounded region, 
i.e., $f$ is closed. In this case the  procedure for defining
$f^{\text{inv}}_{\text{ext}}$ is to take $f^{-1}$ of the usual map $f$ and attribute to $x'=f(x)$
the negative of the same return time $\tau^{\text{inv}}(x')=-\tau(x)<0$ and the inverse reflection coefficient
$R^{\text{inv}}(x')=1/R(x)$. We can therefore write:
\begin{equation}\label{eq.extended-1}
f^{\text{inv}}_{\text{ext}}:\left\{\begin{array}{lll}
\vec{x}_{n+1} & =f^{\text{inv}}(\vec{x}_n) & =f^{-1}(\vec{x}_{n}),  \\
t_{n+1} & = t_n+\tau^{\text{inv}}(\vec{x}_n) & =t_n - \tau(f^{-1}(\vec{x}_{n})),\\
J_{n+1} & = J_n R^{\text{inv}}(\vec{x}_{n}) &= J_n/R(f^{-1}(\vec{x}_n)).
\end{array}
\right.
\end{equation}
 The iteration corresponding to 
(\ref{eq.FP}) of the inverted extended map in cases when $f$ describes a closed map is
\begin{equation}
\begin{array}{ll}  
 \rho_{n+1}(\vec{x}'\equiv{f}^{-1}({\vec{x})}) & =   e^{\kappa^{\text{inv}} \tau^{\text{inv}}(\vec{x})}
                \dfrac{R^{\text{inv}}(\vec{x}) }{\mid \mathcal{D}^{\text{inv}}_f(\vec{x})
                  \mid} \rho_{n}(\vec{x})\\
               &  =   e^{-\kappa^{\text{inv}} \tau(\vec{x}')}
               \dfrac{(1/R(\vec{x}')) }{\mid 1/\mathcal{D}_f(\vec{x}')\mid}\rho_{n}(\vec{x}).
\end{array} 
\label{eq.FP-1}
\end{equation}
This equation is different from the operator of $f_{\text{ext}}$ in
Eq.~(\ref{eq.FP}). Therefore, for extended maps the
chaotic properties ($\kappa$, fractality, etc.) of maps $f_{\text{ext}}$ and
$f_{\text{ext}}^{\text{inv}}$ typically differ. In particular, the rate $\kappa^{\text{inv}}$  of
the inverse map is {\it not}  $-\kappa$ (nor any simple
function of $\kappa$)\footnote{ Even if the dynamics is volume preserving ($\mathcal{D}_f=1$), the
eigenvalue and eigenfunction of Eqs.~(\ref{eq.FP}) and (\ref{eq.FP-1}) differ. If in addition $\tau(x) \equiv 1$, the inverse
map is equivalent to the forward map after $R(\vec{x})$ is replaced by 
$1/R(f^{-1}(\vec{x}))$,  see Eq.~(\ref{eq.FP-1}).}

The difference in the asymptotic rates $\kappa^{\text{inv}} \neq -\kappa$ seems to
contradict the fact that, by definition, maps $f_{\text{ext}}$ and
$f_{\text{ext}}^{\text{inv}}$ cancel each other, i.e.,  the gain and loss resulting
from $n$ applications of one map  is exactly compensated by $n$ applications of
the other map (for arbitrarily large $n$).  The solution 
of this apparent paradox is that the asymptotic densities of the forward $\rho_c$ and
inverse $\rho_c^{\text{inv}}$ maps differ.  Taking $\rho_c$ (or $\rho_c^{\text{inv}}$) as
initial condition $\rho_0$, $\kappa^{\text{inv}}=-\kappa$. However, these are
atypical $\rho_0$. Generic ones converge exponentially fast to the $\rho_c$ of  the
corresponding map.

Many systems of interest are not restricted to a bounded phase space $\Omega$, as
considered above.  For
instance, in scattering systems one usually defines a bounded region $\Gamma$ of interest (which
contains all periodic orbits and the chaotic saddle), but the dynamics is defined in the
full phase space. In this case, the inverse map is defined as above in the whole space,
with the same region of interest $\Gamma$ since the chaotic saddle is invariant.
 Another case of interest  is the one of total absorption in a localized region,
i.e., $R(\vec{x})=0$ for some $\vec{x}\in\Omega$ . In this case, the inverse map can
be defined only as the limiting case of $R \rightarrow 0_+$.

\subsection{Who wins?}

In a general system both gain and absorption coexist and a natural question is whether
decay or explosion is observed globally . This is a priori unclear because
of the non-trivial nature of the asymptotic density $\rho_c(\vec{x})$. When half of the phase space 
has $R_L$ and the other half $R_R$ as reflection
coefficients one would intuitively expect for an ergodic system that a steady state is found with
$R_LR_R=1$. When taking the return times $\tau_i$ into account, intuition says
that for $R_LR_R=1$ the behavior that dominates is the one associated with the half space characterized by
{\em smaller} collision times (with more collisions per time unit). As shown below, 
these two intuitions are not generally correct.

A particularly
interesting case is the one of energetic {\em steady states} when the total energy neither
grows nor decreases in time ($\kappa=0$), in spite of local gains and losses. 
Situations with $\kappa\approx0$ remain unchanged for long times and are thus easy to observe. 
In this case
there is no need for any extraction of intensity and the corresponding iteration scheme is
given by Eq.~(\ref{eq.FP}) without the term $e^{-\kappa \tau(\vec{x})}$, i.e., the
usual Frobenius-Perron equation
for closed systems (even if $f$ is open). 
The density $\rho_c$ is invariant (not conditionally invariant), does
not depend explicitly on the return times~$\tau$,  
and $D_1$ is given by (\ref{eq:D1}) with $\kappa=0$. As shown below, this configuration
leads to a fractal invariant density even for volume-preserving  dynamics.

\section{Examples}\label{sec.examples}

\begin{figure}[bt!]
\includegraphics[width=1\columnwidth]{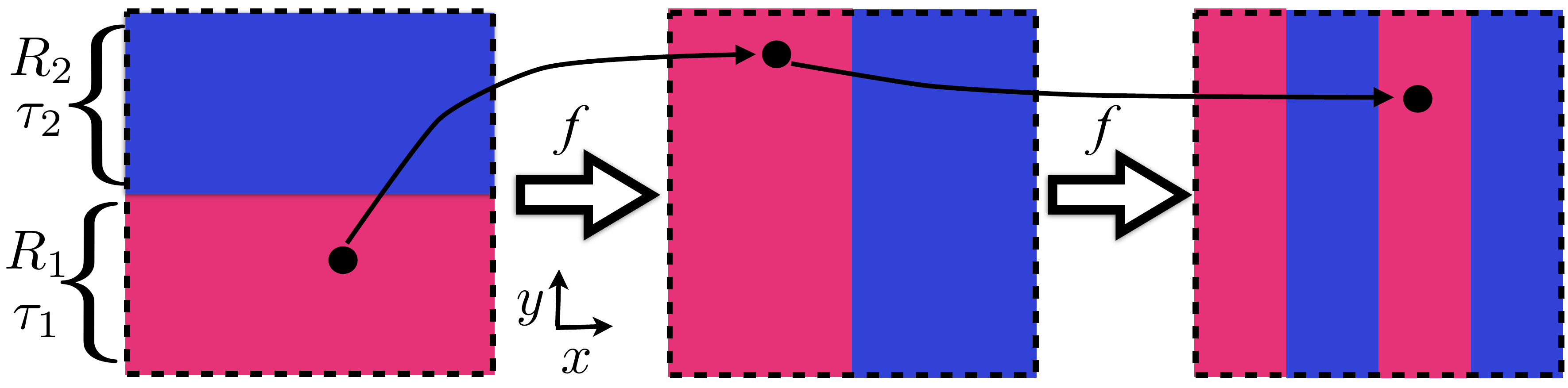} 
\caption{Action of the closed area-preserving baker map $f$ 
with gain on the unit
  square. 
The map $f$ is defined as $(x,y)'=(x/2,2y)$ for $y<0.5$ and $(x,y)'=(x/2+0.5,2y-1)$ for $y>0.5$ and
extended -- as in Eq.~(\ref{eq.extended}) -- by assigning $R=R_1$ ($R=R_2$) and
  $\tau=\tau_1$  ($\tau=\tau_2$) to trajectories in the lower (upper) half.}
\label{fig.baker}
\end{figure}

In this section we use specific systems to explore the
general results reported above.
First, we consider the analytically treatable closed area-preserving baker map defined in
Fig.~\ref{fig.baker}~\cite{Ott-book,LaiTel-book}. Initially $\rho_0 \equiv 1$. In the next step, $\rho_0$ is multiplied
by $R_k e^{-\kappa \tau_k}$, $k=1$ or $2$, leading to two vertical columns of width $1/2$ lying along the $x=0$ and $x=1$ {lines}. 
They carry measures $P_1$ and $P_2$
where
\begin{equation}
P_k=R_k e^{-\kappa \tau_k}/2, \;\;\;\; k=1,2.
\label{eq.P}
\end{equation}
The construction goes on in a self-similar way, 
and the condition that the measure after appropriate compensation remains 
bounded ($\rho_c$ is reached) 
corresponds to $(P_1+P_2)^n=1$, even after $n \gg 1$ steps.
We thus find as an equation specifying the explosion rate $\kappa$:
\begin{equation}
P_1+P_2=1.
\label{eq.r}
\end{equation}   

\subsection{Fractality}

The fractality of the baker map can be calculated explicitly. The generalized
dimensions $D_q^{(2)}$ of $\rho_c$ along the stable (horizontal) direction  is derived
from Eqs.~(\ref{eq.dq}) and~(\ref{eq.P}) as
$D_q^{(2)}=\ln{(P_1^q+P_2^q)}/[(1-q)\ln{2}].$
In the limit of $q \rightarrow 1$ the $D_1$ is obtained
as
\begin{equation} \label{eq.D1}
\begin{array}{ll}
D_1^{(2)} & =  1 - [-\kappa \bar{\tau} +\overline{\ln R}]/\ln 2 \\ 
                 & = 1-\dfrac{-\kappa(\tau_1 P_1 +\tau_2 P_2)+P_1 \ln{R_1} +P_2 \ln{R_2} }{\ln 2},
\end{array}
\end{equation}
which is an equivalent derivation of the general formula~(\ref{eq:D1}) obtained
by identifying $P_i$ with the measure on the chaotic set.

As a particular case, consider $\tau_2=2 \tau_1\equiv 2\tau$ and $R_1=4,
R_2=2$. From~Eq.~(\ref{eq.r}) a quadratic equation is obtained for
$x  \equiv e^{-\kappa  \tau}>0$, leading to $x=\sqrt{2}-1$ and therefore a positive (explosion)
rate $\kappa=-\ln{(\sqrt{2}-1)}/\tau \approx 0.881/\tau$ . Parameters $P_1$ and $P_2$ are
then  $2x$ and $x^2$, respectively. The average return time is 
$\bar{\tau}=\tau 2x + 2 \tau x^2=2 \tau (2-\sqrt{2})$ while
the average of $\ln{R}$ is found to be $ (2\sqrt{2}-1) \ln{2}$.
The numerator of the fraction in the parenthesis of Eq.~(\ref{eq.D1}) is 
$2(2-\sqrt{2})\ln{(\sqrt{2}-1)}+(2\sqrt{2}-1)\ln{2} \approx 0.235$. This is positive, rendering
$D_1^{(2)}=1-0.235/\ln{2} \approx 0.616<D^{(2)}_0=1$, a clear sign of the multifractality of $\rho_c$.

\subsection{Inverse map}\label{sssec.inversebaker}

\begin{figure}[!bt]
\begin{center}
\includegraphics[width=0.85\columnwidth]{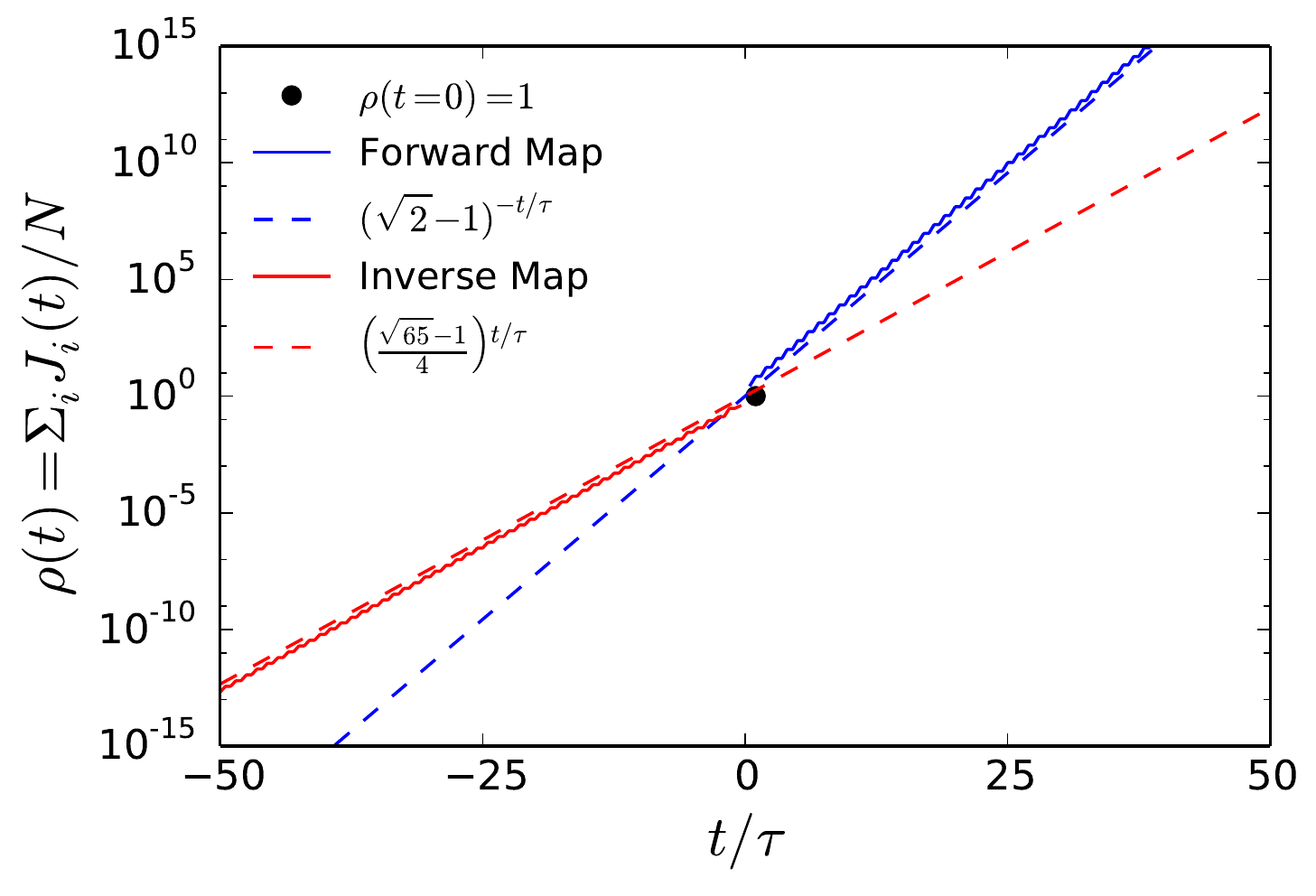}
\end{center}
\caption{ Explosion and escape rates $\kappa$ in the forward and inverse (extended) baker
  map. Points are distributed at $t=0$ uniformly in the phase space with $J_i=1$ and then iterated using 
Eq.~(\ref{eq.extended}) with $f$ defined in Fig.~\ref{fig.baker},
  $\tau_2=2\tau_1\equiv 2\tau$, and $R_1=4=2  R_2$.  The energy density $\rho(t)$ is computed as the average
  intensity of all trajectories $\rho=\sum_{i=1}^{N} J_i/N$ at time $t$.  The forward map
  shows a growing energy in time ($t>0$, 
  blue line) while the corresponding inverse map ($t<0$, red line) shows a decaying
  energy. Dashed lines are the analytical calculations.}
\label{fig.xx}
\end{figure}

The inverse of the baker map  discussed above is computed using the general
relation~(\ref{eq.extended-1}). Symmetries and $\mathcal{D}_f=1$ make
the inverse map  to be equivalent (after a transformation $x\mapsto y, y
\mapsto x$) to the forward map after  replacing $R_i$ by 
$1/R_i$ and $\tau$ by $-\tau$ (see Fig.~\ref{fig.baker}). 
In the example discussed above this
corresponds thus to $\tau_2=2\tau_1=-2\tau, R_1=1/4$,
$R_2=1/2$, which leads to $x \equiv e^{-\kappa \tau} =(\sqrt{65}-1)/4$. This is larger than $1$, implying
a negative $\kappa^{\text{inv}}\approx-0.568/\tau$, i.e. an escape rate, a global decay of
intensity towards zero.   In Fig.~\ref{fig.xx} we
confirm that these analytical calculations agree with simulations of individual
trajectories. The different slopes confirm the general result $\mid\kappa\mid \ne \mid \kappa^{\text{inv}}\mid$. In the previous section we
related this apparent paradox to the difference between $\rho_c$ and
$\rho_c^{\text{inv}}$, which can be quantified by $D_1$. From Eq.~(\ref{eq.D1}), $D_1^{(2)
  \text{inv}}\approx0.783\neq  0.66 1\approx D_1^{(2)}$. Non-trivial fractality is preserved despite the change of sign in $\kappa$.

\subsection{Who wins?}

We allow for competition between gain and absorption in the particular example defined above by writing $R_i(\alpha)=\alpha R_i$, with
$0<\alpha<1$.
Because of (\ref{eq.r}), this leads with $R_1=4=2 R_2$ to an $\alpha$-dependent explosion rate $\kappa(\alpha)$
given by  $e^{-\kappa(\alpha) \tau}=\sqrt{1+1/\alpha}-1$.
Decreasing $\alpha$ from unity, this quantity is less than unity but increases
with decreasing $\alpha$. At $\alpha \lessapprox 1/2$ absorption and gain coexist ($R_2(\alpha)<1$)
but explosion still dominates ($\kappa>0$).   For $\alpha < \alpha_c=1/3$ absorption
dominates ($\kappa<0$).   At the critical value $\alpha=\alpha_c$, $\kappa=0$, which is the steady state
condition\footnote{ For general $R_i$ and $\tau_i$,  Eq.~(\ref{eq.r}) yields 
$\alpha_c=2/(R_1+R_2)$.},  $R_{1c} \equiv R_1(\alpha=\alpha_c)=4/3$, $R_{2c}\equiv
R_2(\alpha=\alpha_c)=2/3$, $P_{1c}=2/3$,  $P_{2c}=1/3$, 
the average of $\ln{R}$ is $5/3 \ln{2}- \ln{3}$, and thus Eq.~(\ref{eq:D1}) yields 
again fractality $D_1^{(2)} =1 -\overline{\ln{R}}/\ln{2}=0.919$ for any $\tau_1, \tau_2$. 

Even  though the return times $\tau_i$ are irrelevant at the steady state,
the steady state is not achieved at $R_1R_2=1$.  In our example, $R_1(\alpha)R_2(\alpha)=8 \alpha^2=1$ for 
$\alpha=1/\sqrt{8}= 0.354>\alpha_c=1/3$ (explosion). This remains valid in maps ($\tau_i=1$) for any $R_1,R_2$:
introducing $R_1=1/R_2$ in Eq.~(\ref{eq.r}) leads to  $\kappa
\tau=\ln{((R_1+1/R_1)/2)}\ge0$. 
The stead-state condition is 
thus not $R_1R_2=1$, but rather $R_1+R_2=2$.   

\begin{figure}[!bt]
\begin{center}
\includegraphics[width=0.95\columnwidth]{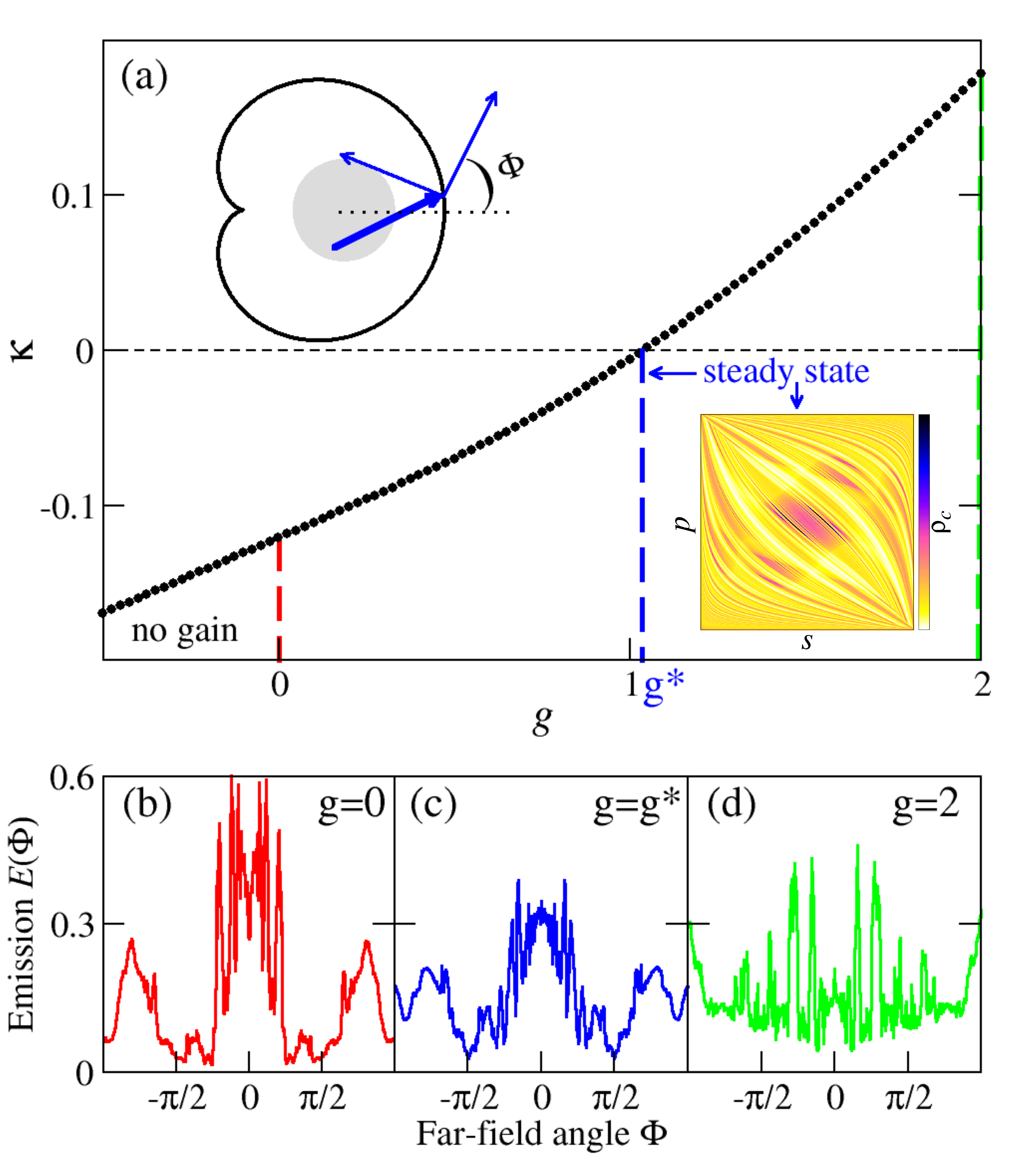}
\end{center}
\caption{Ray emission from an optical cardioid billiard with gain and transmission.  (a)
  Asymptotic rate~$\kappa$ as a function of the gain strength $g$. The steady state
  $\kappa=0$ is  achieved at $g=g^* \approx 1.0367$, the lower inset shows the invariant density
    $\rho_c$ for this case. (b-d) Long-time ($t\gg1$) emission
      $E(\Phi)$ of intensity $J$ in the far-field angle $\Phi$ (normalized as $\int E(\Phi) d\Phi =1$) for (b) $g=0$, 
no gain and decaying intensity: $\kappa <0$; (c) $g=g^*$, steady state, $\kappa=0$; (d) $g=2$, explosion, $\kappa>0$. The intensity of rays grows with rate $g$ at the gain
  region (gray circle, see also Fig.~\ref{fig1}) and is split between reflected and
  transmitted rays for small collision angles $\theta$. This leads to a partial optical reflection
  $R_o(\theta)=[\sin(\theta_T-\theta)/\sin(\theta_T+\theta)]^2<1$ for $\mid p \mid \equiv
  \mid \sin  \theta \mid <1/\eta$, where we used $\eta=3$ and the transmitted angle $\theta_T$ is given by Snell's law as $\eta \sin \theta =\sin
  \theta_T$~\cite{Lee2004,Ryu2006}.}
\label{fig.optical}
\end{figure}

As a final illustration of the significance of our general results, we perform numerical
simulation of rays in an optical cavity with gain and absorption. Imagine
that the cardioid billiard with localized gain introduced in Fig.~\ref{fig1} is composed
of a dielectric material (with refraction index $\eta$). 
Figure~\ref{fig.optical} reports results for transverse-electric polarized light in such a configuration.
Whether explosion occurs, depends on the gain parameter $g$ ($\kappa$
depends smoothly on $g$, see panel a). At the critical value $g=g*$ gain
and loss cancel each other ($\kappa=0$) and an energetic steady state sets in. The
density~$\rho_c$ is fractal for any $g$, also at $g=g^*$  
(see lower inset of panel a). 
 The transmitted rays can be detected outside the billiard as an emission in a given direction (represented by
the far-field angle $\Phi$, see  upper inset in
Fig.~\ref{fig.optical}a).  We obtain that the (observable) far-field emission (as a
function of angle $\Phi$) is modified by the gain (compare panels b-d). We thus conclude
that (non-uniform\footnote{For gain uniform in the cavity  ($\tau_g=\tau$),
  Eq.~(\ref{eq.FP}) shows that $\kappa$ is re-scaled
but $\rho_c$ (and therefore the emission) is not changed.}) gain has to be included in ray simulations.

\section{Conclusions}\label{sec.conclusions}

We considered chaotic systems in which the intensities of trajectories may grow in time
(e.g., due to gain or a reflection coefficient $R>1$). We extended the formalism of
systems with absorption ($R<1$) to show how the theory of (open) chaotic systems can be
used in this new class of systems. For instance,  we derived a formula --
Eq.~(\ref{eq:D1}) -- that  relates the invariant properties (e.g., fractal 
dimensions and the explosion/escape  rates) of systems with gain/absorption,  a
generalization of important results in the theory of open systems (in which $R$ is
either $0$ or $1$)~\cite{GaspardBook,LaiTel-book}. Despite the unifying formalism, our 
results reveal an 
intricate relationship between systems with gain and with absorption. For instance, the
inverse of a system with gain is a system with absorption, but -- in contrast to usual
dynamical systems -- their invariant properties are not
trivially related to each other. 

For applications in optical microcavities, our results show how gain can be incorporated
in the usual ray simulations. Whenever the  gain is not uniform in the cavity,  we find
that the emission is modified and can be described through our formalism of chaotic
explosions. These results can be tested experimentally by constructing optical cavities
with different localized gain regions~\cite{Shinohara2011} and comparing the
emission to ray simulations with and without gain.

\section{Acknowledgments} We thank M. Sch\"onwetter for the careful reading of the manuscript.
This work has been supported by OTKA   grant No. NK100296 and by the Alexander von
Humboldt Foundation.

\end{document}